\documentclass[twocolumn,english,aps,floatfix,superscriptaddress,amsfonts,amssymb,superscriptaddress]{revtex4}
\usepackage{color}

\usepackage{graphicx}
\usepackage{amsmath}
\usepackage{latexsym}
\usepackage{epstopdf}
\usepackage{amsmath}
\usepackage{amssymb,amsmath}
\usepackage{multirow}
\usepackage{mathtools}
\newcommand{\beq}{\begin{equation}}
\newcommand{\eeq}{\end{equation}}

\usepackage{lmodern} \usepackage[T1]{fontenc}
\usepackage{tikz}
\usepackage{xcolor}
\usepackage{subcaption}
\usetikzlibrary{patterns}

\begin{document}

\title{ On the possibility of complete revivals after quantum quenches to a
    critical point}
\author{K. Najafi}
\affiliation{ Department of Physics, Georgetown University, 37th and O Sts.  NW, Washington, DC 20057, USA}
\author{M.~A.~Rajabpour}
\affiliation{  Instituto de F\'isica, Universidade Federal Fluminense, Av. Gal. Milton Tavares de Souza s/n, Gragoat\'a, 24210-346, Niter\'oi, RJ, Brazil}

\date{\today{}}

\begin{abstract}

In a recent letter, J. Cardy, Phys. Rev. Lett. \textbf{112}, 220401 (2014), the author made a very interesting observation that complete revivals of quantum states after quantum quench can happen in a period which is a fraction of the system size. This is possible for critical systems that can be described by minimal conformal field theories (CFT) with central charge $c<1$. 
In this article, we show that these complete revivals are impossible in microscopic realizations of those minimal models. We will prove the absence of the mentioned complete revivals for the critical transverse field Ising chain analytically, and present numerical results for the critical line of the XY chain. In particular, for the considered initial states, we will show that criticality has no significant effect in partial revivals. We also comment on the applicability of quasi-particle picture to determine the period of the partial revivals qualitatively. In particular, we detect a regime in the phase diagram of the XY chain which one can not determine the period of the partial revivals using the quasi-particle picture.

\end{abstract}
\maketitle
\section{Introduction}

Quantum mechanical version of Poincar\'e recurrence theorem guarantees that any system with discrete energy eigenstates, after a sufficiently long  but finite time, will return to a state which is very close to it's initial state. Although this seems a very natural expectation
for many body systems, it usually takes astronomical times to see an (almost) complete revivals. However, in some systems, partial revivals
are possible which makes the problem a very interesting subject, see for example\cite{Quan2006,Yuan2007,Rossini2007,Zhong2011,Happola,Montes2012,Sharma2012,Rajak2014,Jafari2017,Igloi2011,Zhang2009,Sacramento2006,DS2011}.
The problem of revivals or related phenomena also appears in many other concepts such as dynamical transition and
quantum speed limit \cite{Heyl2013,Pollmann2010,Karrasch2013,Heyl2017}. Quite naturally, one usually is interested in the problem of revivals
when the number of particles is limited or in other words when there is a finite size effect. The presence of the finite size effect
usually makes the exact calculations difficult, however, it has the benefit of being accessible by the numerical means.

In a recent letter \cite{Cardy2014}, the author has made a very interesting analytical calculation by using conformal field theory for finite system size and observed
that the complete revivals are possible for Loschmidt amplitude in critical systems in a period which is a fraction of the system size. This is quite unexpected because any
full revival requires a nearly perfect fine-tuning of phase conditions. To the best of our knowledge, this is the first example of the prediction of
full revivals in many-body quantum systems in an accessible time.
In this brief article, we make a closer look to this phenomena in microscopic systems.
In particular, we show that the  full revivals of \cite{Cardy2014} are impossible in
microscopic realizations of those conformal field theories that studied in \cite{Cardy2014}. 
In the next section, first, we briefly review the arguments 
in favor of and against the presence of complete revivals in critical systems. Then in section three, we study the Loschmidt echo (fidelity) in quantum XY chain and find an exact determinant formula for a particular initial state. In section four, we calculate the fidelity at the critical point of the periodic (open) transverse field Ising chain analytically (numerically), and show that the complete revivals discussed in \cite{Cardy2014} are absent. In section five, we explore the other parts of the phase diagram of the XY chain. In particular, we study the quasi-particle picture for different post-quench Hamiltonians and show that the picture can determine the periods of the revivals just in some part of the phase diagram. Finally, in the last section, we conclude our paper.

\section{Revivals in conformal field theories}

Consider a one-dimensional periodic quantum chain of length $L$ and the Hamiltonian $H$.
Assume an initial state which is very close to a conformally invariant state $|B\rangle$. Since the conformal states are non-normalizable, one needs to introduce a parameter $\beta$ which is called extrapolation length and then one can write the initial state as $|\psi_0\rangle\approx e^{-\frac{\beta}{4} H}|B\rangle$.
The extrapolation length is usually of the order of a few lattice sites, in other words, we have $L\gg\beta$.
The parameter $\beta$ can be estimated 
 by calculating the expectation value of the Hamiltonian as $\langle\psi_0|H|\psi_0\rangle=\frac{\pi cL}{6\beta^2}$, see \cite{Cardy2014}.
To show the complete revival, the reference \cite{Cardy2014} calculates the fidelity defined as
\begin{eqnarray}\label{Fidelity}
F(t)=|\langle\psi_0|e^{-iHt}|\psi_0\rangle|
\end{eqnarray}
using the well-established recipe, see for example
 \cite{DS2011}. Based on his argument, for minimal models at large $\frac{L}{\beta}$, there must be always complete revivals at multiples of $t=M\frac{L}{2}$, where $M\sim\frac{24}{1-c}$.
Not surprisingly in the regime $L\gg\beta$, there is no effect of the extrapolation length in the revival times. Although there is nothing wrong in the CFT calculations in \cite{Cardy2014}, this effect can not be seen in a microscopic quantum chain. There are two good reasons:
The first reason, which is already noticed in \cite{DS2011}, is related to the presence of the excited states in any global quantum quench which in principle
can not be described by CFT. The second reason is that \cite{Cardy2014} assumes that there is a one to one correspondence
between an initial state in a microscopic system and conformal boundary states which in general is not true. There are many discrete
initial states, probably exponentially growing with the system size, that flow to the same conformal boundary states either with the same $\beta$ or different extrapolation lengths. 
This means that although in a CFT setup the system comes back to itself with probability one, in the discrete model it can be in a state which is completely different but still with the same continuum description. Although  in principle, this problem can
be resolved by considering all the possible irrelevant perturbations of the CFT, see for example \cite{Cardy2014, Cardy2016}  it will eventually affect the complete revivals anyway.
The reference \cite{Cardy2014} discusses, in particular, a quench from the disordered phase in the transverse field Ising chain and shows that there should be complete
revivals at times $t=\frac{nL}{2}$ for even $n$, while for odd $n$ the complete revivals are suppressed. In the next section, we will show that
the complete revivals are absent in the critical transverse field Ising chain.

\section{Loschmidt echo in quantum XY chain}

In this section, we study the revivals in the quantum spin chain when the initial state is the case with all spins $\sigma^z$ are up or down.
The Hamiltonian of XY-chain is as follows:
\begin{eqnarray}\label{HXY1}\
H_{XY}=\hspace{7cm}\\ \nonumber-J\sum_{j=1}^{L}\Big{[}(\frac{1+a}{2})\sigma_{j}^{x}\sigma_{j+1}^{x}+(\frac{1-a}{2})\sigma_{j}^{y}\sigma_{j+1}^{y}\Big{]}-h\sum_{j=1}^{L}\sigma_{j}^{z}.
\end{eqnarray} 

Different phases of the model for $J=1$ are shown in the Figure 1. The line $a=1$ is the 
transverse field Ising chain. The  $h=1$ line is critical for all the values of $a$ and we call it XY critical line.
On the circle $a^2+h^2=1$, the wave function
of the ground state is factorized into a product of single spin states \cite{Kurmann1982}.

 \begin{figure} [hthp!] \label{fig1}
\includegraphics[width=0.45\textwidth,angle =-00]{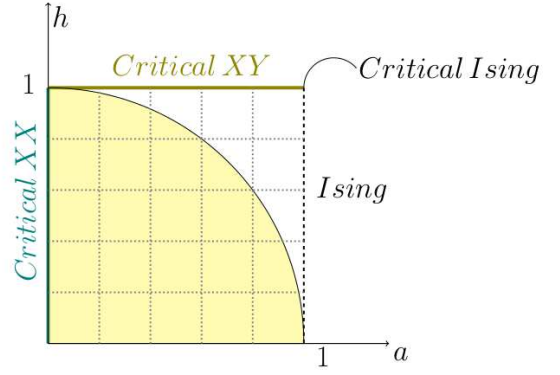}
\caption{Color online) Different  regions in the phase diagram of the quantum $XY$ chain. The critical $XX$ 
chain has central charge $c=1$ and critical $XY$ line has $c=\frac{1}{2}$. The region $a^2+h^2<1$ is depicted with the yellow color. } 
\label{fig:XYpahse space}
\end{figure}

After using the Jordan-Wigner transformation 
$c_j^{\dagger}=\prod_{l<j}\sigma_l^z\sigma_j^{+}$, which maps the Hilbert space of a quantum chain of a spin $1/2$ into the Fock space of  
spinless fermions, the new Hamiltonian becomes 
\begin{eqnarray}\label{XY-ff}
H=J\sum_{j=1}^{L-1} (c_j^{\dagger}c_{j+1}+ac_j^{\dagger}c_{j+1}^{\dagger}+h.c.)-\sum_{j=1}^{L}h(2c_j^{\dagger}c_j-1)\\ \nonumber
+\mathcal{N}J(c_L^{\dagger}c_1+ac_L^{\dagger}c_1^{\dagger}+h.c.),\hspace{3cm}
\end{eqnarray}
where $c_{L+1}^{\dagger}=0$ and $c_{L+1}^{\dagger}=\mathcal{N}c_{1}^{\dagger}$ for open and periodic boundary conditions respectively with $\mathcal{N}=\prod_{j=1}^{L}\sigma_j^z=\pm1$.
The above Hamiltonian can be written as:
\begin{eqnarray}\label{H1}\
\textbf{H}=\textbf{c}^{\dagger}.\textbf{A}.\textbf{c}+\frac{1}{2}\textbf{c}^{\dagger}.\textbf{B}.\textbf{c}^{\dagger}+\frac{1}{2}\textbf{c}.\textbf{B}^{T}.\textbf{c}-\frac{1}{2}{\rm Tr}{\textbf{A}},
\end{eqnarray}
with appropriate $\textbf{A}$ and $\textbf{B}$ matrices as:
\begin{eqnarray}\label{mat_A}\
\textbf{A}= \begin{pmatrix}
     -2h     & J     & 0      &\dots           &\mathcal{N}J \\
     J       &-2h    &J       &0               &0           \\  
     0       & J     &-2h     &J             &0           \\       
    \ddots      &\ddots      & \ddots &\ddots    &\ddots               \\  
\mathcal{N}J  &0            &\dots &J & -2h  \\ 
  \end{pmatrix}  \nonumber, 
\end{eqnarray} 
\begin{eqnarray}\label{mat_B}\
\textbf{B}= \begin{pmatrix}
     0       & aJ     & 0      &\dots        &-aJ\mathcal{N} \\
     -aJ     &0      &aJ           & &0                \\  
     0       & -aJ    &0       &aJ    & 0                 \\       
     \ddots       &\ddots      & \ddots &\ddots    &\ddots           \\ 
aJ\mathcal{N} &0   & \dots&-aJ &0 \\ 
  \end{pmatrix}.
\end{eqnarray}  
To calculate the Loschmidt echo, first we decompose $e^{-i\textbf{H}t}$ using the Balian-Brezin formula \cite{Balian1969} as:
\begin{eqnarray}\label{H2}\
e^{-i\textbf{H}t}=e^{\frac{1}{2}\textbf{c}^{\dagger}\textbf{X}\textbf{c}^{\dagger}}e^{\textbf{c}^{\dagger}\textbf{Y}\textbf{c}}e^{-\frac{1}{2}{\rm Tr}\textbf{Y}}e^{\frac{1}{2}\textbf{c}\textbf{Z}\textbf{c}},
\end{eqnarray} 
where $\textbf{X}$, $\textbf{Y}$, $\textbf{Z}$ can be calculated from the blocks of matrix
$\textbf{T}$ defined as 
\begin{eqnarray}\label{mat_T}\
\textbf{T}=e^{-it \begin{pmatrix}
\textbf{A} & \textbf{B}\\
\textbf{-B} & \textbf{-A}\\
  \end{pmatrix}  }
 = \begin{pmatrix}
\textbf{T}_{11} & \textbf{T}_{12}\\
\textbf{T}_{21} & \textbf{T}_{22}\\
  \end{pmatrix},  
\end{eqnarray} 
Then we have
\begin{eqnarray}\label{X}\
\textbf{X}=\textbf{T}_{12}(\textbf{T}_{22}^{-1}),\hspace{0.5cm}\textbf{Z}=(\textbf{T}_{22}^{-1})\textbf{T}_{21},\hspace{0.5cm}e^{\textbf{-Y}}=\textbf{T}_{22}^{T}.
\end{eqnarray} 
Note that we always have $\textbf{Z}=-\textbf{X}$. Finally, the fidelity for the desired initial state (all the spins up) will be: 
\begin{equation} \label{fidelity}
F(t)=|\langle0|e^{-i\textbf{H}t}|0\rangle|=|\det(\textbf{T}_{22})|^{\frac{1}{2}}.
\end{equation}
The same formula is also valid for the case when the initial state is all the spins down, in other words when all the sites are filled with fermions $|1\rangle$.
In the next subsections, we will use the above formula to study the revivals in different phases of the quantum spin chain. Note that although we will keep the $J$ coupling explicitly in some of the formulas for numerical calculations we always take $J=1$.

\section{Revivals in critical transverse field Ising chain}

In this section, first we calculate the fidelity for the periodic critical Ising point exactly, then we will study
the open chain numerically.
\subsection{Periodic critical transverse field Ising chain}
Consider a periodic critical Ising model Hamiltonian in Ramond sector after Jordan-Wigner transformation, equation (\ref{XY-ff})
with $\mathcal{N}=J=a=h=1$. Since in this case the matrices $\textbf{A}$ and $\textbf{B}$ commute, one can calculate the fidelity exactly. Note that
in this case
 the matrix $\begin{pmatrix}
\textbf{A} & \textbf{B}\\
\textbf{-B} & \textbf{-A}\\
  \end{pmatrix}$ is a circulant matrix which gaurranties the exact calculation of the eigenvalues of the matrix $\textbf{T}_{22}$ with classical methods.
After expanding (\ref{mat_T}) we have
\begin{eqnarray}\label{Texp6}\
\textbf{T}_{22}&=&\textbf{T}_{11}^*=\cosh(2t\sqrt{\textbf{A}})+\frac{i\sqrt{A}}{2}\sinh(2t\sqrt{\textbf{A}}),\\
\textbf{T}_{12}&=&-\textbf{T}_{21}=-it\textbf{B}[\frac{\sinh(2t\sqrt{A})}{2t\sqrt{A}}].
\end{eqnarray} 
Although it is not needed for our future discussion, we also report the exact form of the 
matrices $\textbf{X}$:
\begin{eqnarray}\label{X1}\
\text{X}=\frac{-i\textbf{B}}{2{\sqrt{\textbf{A}}\coth[2t\sqrt{\textbf{A}}]}+i\textbf{A}},
\end{eqnarray} 
Since the eigenvalues of the matrix $\textbf{A}$ are $\lambda_j=-2+2\cos\frac{2\pi j}{L}$, where $j=1,2,...,L$ ;
logarithmic fidelity for the critical periodic Ising chain can be written explicitly as
\begin{equation} \label{fidelity Ising}
\ln[ F(t)]=\frac{1}{4}\sum_{j=0}^{L-1}\ln[1-\cos^{2}[\frac{\pi j}{L}]\sin^2[4t\sin[\frac{\pi j}{L}]])].
\end{equation}
In Figure 2, one can see that although there are partial revivals at multiples of $t=\frac{L}{4}$, which can be understood
with the quasi-particle picture, the complete revivals do not happen. Of course if one waits enough time, there will be always almost complete revivals but they are usually expected to happen in much larger times that are usually inaccessible. Note that for the considered  initial state, we have
 $\langle\psi_0|H|\psi_0\rangle=L$ which means that $\beta=\sqrt{\frac{\pi}{12}}$ or in other words we are in a regime that $\frac{L}{\beta}$
 is very large which is well inside the regime considered in \cite{Cardy2014}. 

\begin{figure} [hthp!] \label{fig2}
\centering
\includegraphics[width=0.40\textwidth,angle =-00]{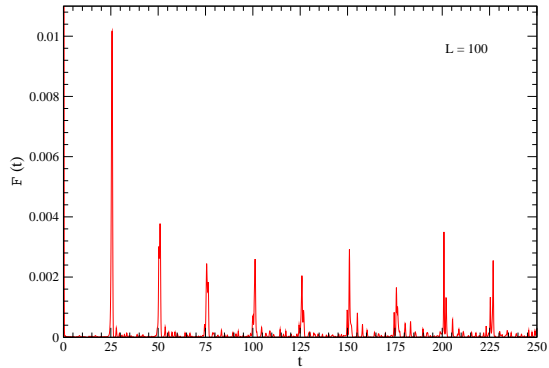}
\caption{(Color online) logarithmic fidelity for the periodic critical Ising chain starting from the all aligned spins $\sigma^z$ initial state. } 
\end{figure}

\subsection{Open critical transverse field Ising chain}
In this subsection, we repeat the analyses of the previous section for the open chains to see the effect of the boundary condition on
the revivals. Unfortunately, we were not able to provide an exact result in this case so the calculations are based on a numerical evaluation of the determinant in the equation (\ref{fidelity}). The main reason for our failure   at the critical Ising point can be traced back to this fact that in this case the two matrices $\textbf{A}$ and $\textbf{B}$
do not commute and so the expansion method gets too complicated after few steps. Also note that in this case the matrix $\begin{pmatrix}
\textbf{A} & \textbf{B}\\
\textbf{-B} & \textbf{-A}\\
  \end{pmatrix}$
is not a  circulant matrix and so the common methods of diagonalization can  not be applied. The numerical results depicted in Figure 3 confirms the absence
of the  complete revivals introduced in \cite{Cardy2014} and the usefulness of the quasi-particle picture. We will come back to a more detailed study of the quasi-particle picture in the next section.
 \begin{figure} [hthp!] \label{fig3}
\includegraphics[width=0.40\textwidth,angle =-00]{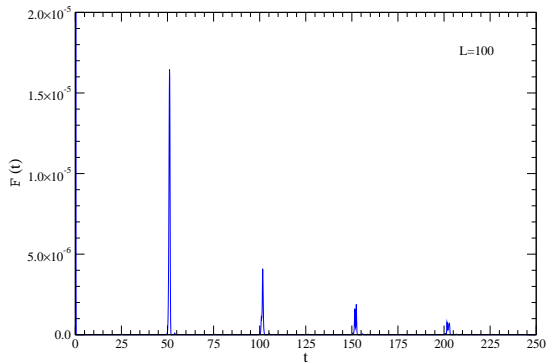}
\caption{(Color online) logarithmic fidelity for the open critical Ising chain starting from the all aligned spins $\sigma^z$ initial state. } 
\end{figure}
\section{Revivals and quasi-particle picture}
In this section, we extend the analyses of the previous section to the other parts of the phase diagram of the XY chain. In addition, we also examine the applicability of the quasi-particle picture in determining the periods of the revivals in the Loschmidt echo.
First, we make a brief comment on the quasi-particle picture, see \cite{CC2006}. Based on this semi-classical picture the pre-quench state has more energy than the
post-quench Hamiltonian ground state and so consequently, the initial state plays the role of a source of quasi-particles. The quasi-particles with the maximum group velocity usually
are the ones that can be connected to the saturation of the entanglement entropy \cite{CC2007} or the revivals in the Loschmidt echo \cite{DS2011}. The
dispersion relation and the group velocity  of the Hamiltonian (\ref{XY-ff}) are
\begin{eqnarray}\label{dispersion}\
\epsilon_k&=&J\sqrt{(\cos\phi_k-h)^2+a^2\sin^2\phi_k},\\
v_{g}&=&J\sin\phi_k\frac{2a^2\cos\phi_k-\cos\phi_k+h}{\sqrt{(\cos\phi_k-h)^2+a^2\sin^2\phi_k}}.
\end{eqnarray} 
where $\phi_k=\frac{2\pi}{L}(k+\frac{\mathcal{N}-1}{4})$ with $k=0,...,L-1$. In Figure 4, we depicted $v_{g}$ for 
different values of $a$ and $h$. Note that for sufficiently large $L$, there is no significant difference between open and periodic cases.
 \begin{figure} [hthp!] \label{fig4}
\includegraphics[width=0.45\textwidth,angle =-00]{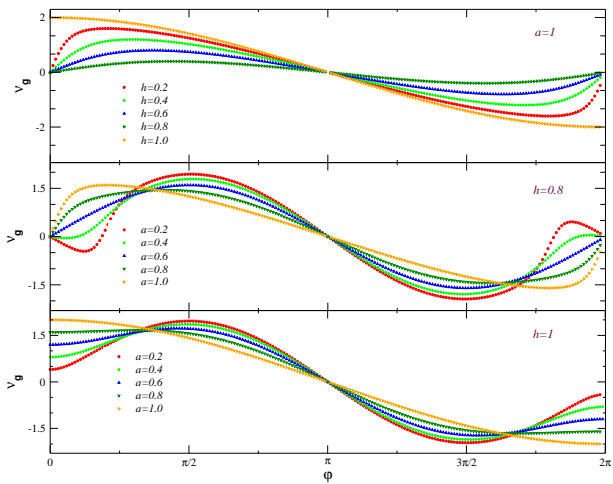}
\caption{(Color online) Group velocity $v_{\phi}$ with respect to $\phi$ for different values of $a$ and $h$.} 
\end{figure}
 \begin{figure} [hthp!] \label{fig5}
\includegraphics[width=0.45\textwidth,angle =-00]{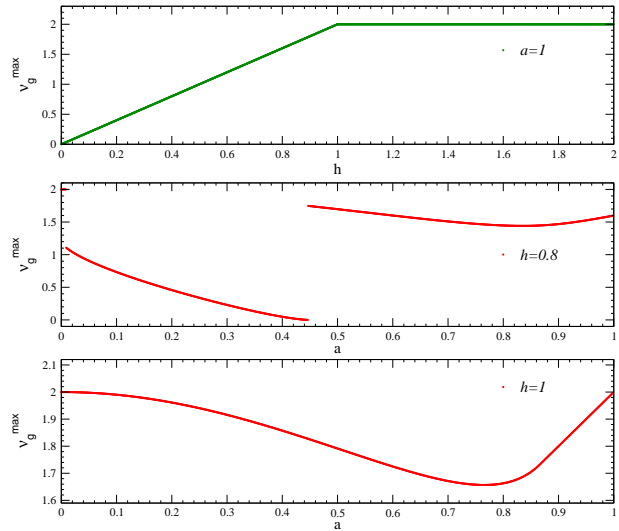}
\caption{(Color online) Maximum group velocity for different values of $a$ and $h$. } 
\end{figure}
Using the above formula one can derive the maximum group velocity $v_g^{max}$ for different values of $a$ and $h$, see Figure 5.
Having these group velocities for quasi-particles, one can guess the following periods for the appearance of revivals
in the Periodic and Open quantum chains:
\begin{eqnarray}\label{time revivals}\
T_{p}=\frac{L}{2v_{g}}, \hspace{1cm} T_{o}=\frac{L}{v_{g}},
\end{eqnarray} 
where  $v_{g}$ is usually the velocity of the fastest quasi-particles $v_g^{max}$. However, this is not a rule and sometimes other quasi-particles 
can carry more information than the fastest quasi-particles as it was discussed in the context of entanglement entropy in \cite{Fagotti2011}
and in the context of Loschmidt echo in \cite{DS2011}. In those cases $v_{g}$ in the equation (\ref{time revivals})
will be different from $v_g^{max}$. We are not aware of a criterion which one can use apriory to decide what is the most important group velocity.
In the next three subsections, we study the revivals and the quasi-particle picture in different regimes.

\subsection{Ising line: $a=1$}
In Figure 6, we depicted the Loschmidt echo for different values of $h$. Two comments are in order: first of all, 
 at non-critical points similar to the critical point we have partial revivals. Apart from the period of the revivals, there is no significant difference in
the form of the Loschmidt echo at and outside of the critical point. Secondly, the period of the revivals can be understood by 
 taking the maximum group velocity $v_g^{max}=2h$ as the relevant velocity.
  \begin{figure} [hthp!]
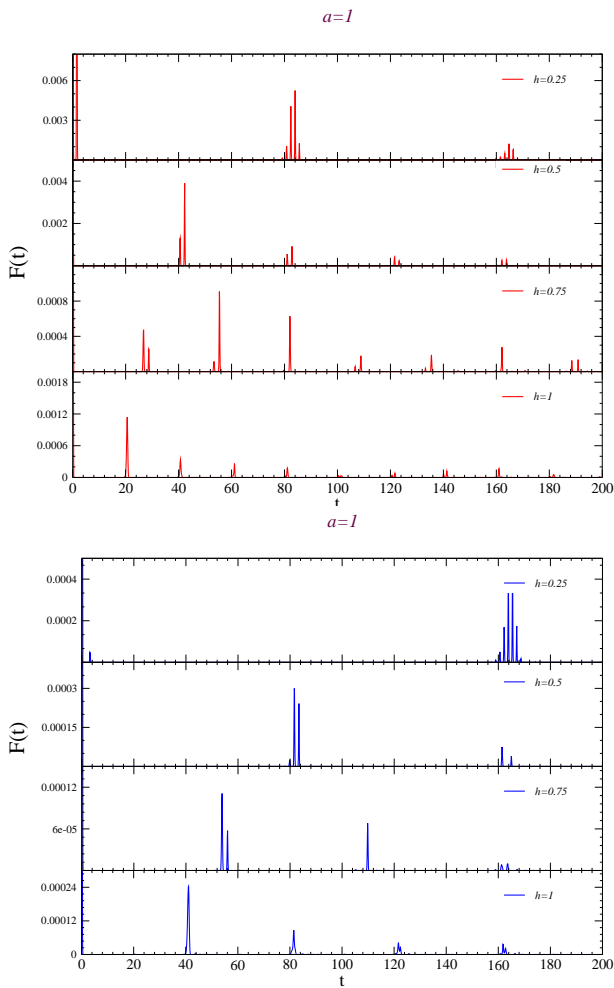
 \label{fig6}
\includegraphics[width=0.45\textwidth,angle =-00]{fig6a.eps}
\includegraphics[width=0.45\textwidth,angle =-00]{fig6b.eps}
\caption{(Color online) Loschmidt echo on the Ising line for periodic and open chains in top and bottom panel  respectively($L=80$). The revivals are compatible with quasi-particle picture with maximum group velocity $v_g^{max}=2h$.}
\end{figure}
 \subsection{Critical XY line: $h=1$}
 This is a critical line which it is in the same universality class as the critical Ising chain.
 The results for the Loschmidt echo on different points are shown in Figure 7.
 The interesting fact is that in this case, the relevant group velocity is clearly $|v_g^f|=2a$ which is the Fermi velocity.
 As it was already discussed in the context of the Loschmidt echo after local quenches in \cite{DS2011}, it
 is not the maximum group velocity for $a<\frac{\sqrt{3}}{2}$. This is an interesting example
of a case which the relevant group velocity is different from the maximum group velocity.
  \begin{figure} [hthp!]
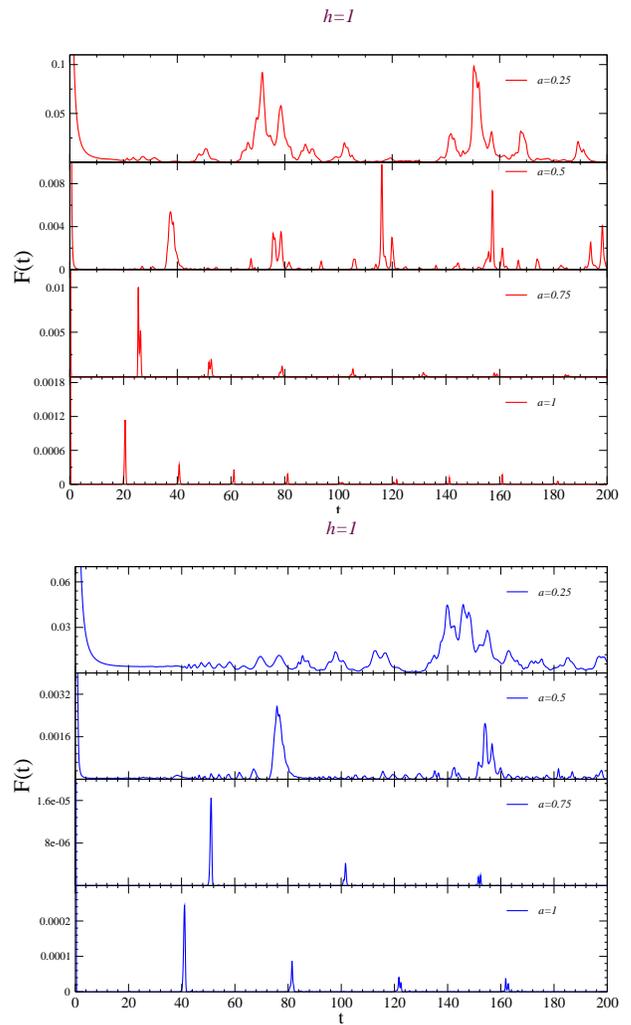
 \label{fig7}
\includegraphics[width=0.45\textwidth,angle =-00]{fig7a.eps}
\includegraphics[width=0.45\textwidth,angle =-00]{fig7b.eps}
\caption{(Color online) Loschmidt echo on the XY critical line for periodic and open chains in top and bottom panel  respectively($L=80$). The revivals are compatible with quasi-particle picture
with the Fermi group velocity $v_g^f=2a$.} 
\end{figure}
 \subsection{Non-critical regime: $0<h,a<1$}
Based on the results of the above two subsections one might be tempted to guess that since this region is non-critical, the quasi-particle picture with the maximum group velocity should be appropriate to guess the form of the partial revivals. However, strikingly as one can see in Figure 8,
  there are two completely different regimes with very different behaviors. In the region $a^2+h^2>1$
 the quasi-particle picture with the maximum group velocity works perfectly, however, in the region $a^2+h^2\leq1$
 there seems to be no clean way to attribute the revivals to quasi-particles with fixed velocities.
 One might understand it as a regime that there are more than one type of important quasi-particles which their velocity difference kill clean periodic revivals. It is absolutely unclear why the line $a^2+h^2=1$ should separate these two regimes. To keep the figure simple, we have only depicted four different points, however we have checked many other different points and confirmed numerically that indeed this line is  at the border between the two different regimes.
%
%

 \section{Conclusions}
  In this paper, we studied revivals in the XY chain starting from an initial state with all the spins $\sigma^z$ in the direction of the transverse field. Our   conclusions are the followings: 
first of all, we proved that complete revivals in the times introduced by \cite{Cardy2014}, cannot happen in the microscopic quantum critical chains.
Secondly, for the considered initial state we showed that there are three interesting regimes. For $a^2+h^2>1$, one can use
the quasi-particle picture to predict the period of the partial revivals. On the critical 
XY line $h=1$, one must use the Fermi velocity $v_g^f=2a$ to calculate the revivals. However, for the other points, the maximum group velocity $v_g^{max}$ is the important group velocity. For the region $a^2+h^2\leq1$,
the revivals do not follow a clean periodic structure which indicates the presence of more than one type of important quasi-particles.
It will be interesting to study the structure of revivals in also other models, especially interacting models such as the XXZ chain. Loschmidt echo
in the non-critical phase of the XXZ chain 
has been already studied  in \cite{XXZ1,XXZ2}, however, it seems that the problem of revivals in the critical regimes of interacting models has not been studied in full detail sofar.
In particular, it is very important to study the effect of the initial state in these models.

\begin{widetext}

\begin{figure}[hthp!]
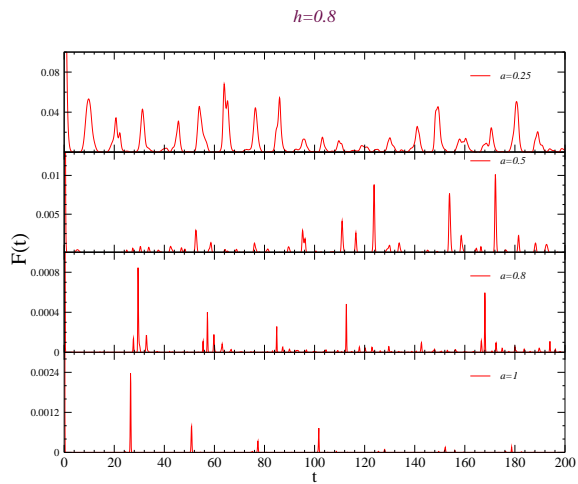

    \centering
    \begin{subfigure}{\textwidth}
        \centering
        \includegraphics[width=0.42\textwidth,angle =-00]{fig8a.eps}
    \end{subfigure}%
    ~ 
    \begin{subfigure}{\textwidth}
        \centering
        \includegraphics[width=0.42\textwidth,angle =-00]{fig8b.eps}
    \end{subfigure}
    \caption{(Color online) Loschmidt echo on the line $h=0.8$ for periodic and open chains in left and right panel respectively($L=80$). The revivals are compatible with quasi-particle picture with maximum group velocity (see Figure 5) as far as $a^2+h^2>1$. For the region $a^2+h^2\leq 1$ there are no clean periodic revivals.}
\end{figure}

\end{widetext}

\textbf{Acknowledgment:}
 We are grateful to J. Dubail for taking our attention to the argument in \cite{DS2011} and useful comments.
 The work of MAR is supported partially by CNPq, Brazilian agency. The work of KN is supported by 
 National Science Foundation under grant number PHY- 1314295.

\end{document}